\documentclass{iau}
\usepackage{graphicx,natbib}
 
\newcommand{\apj}{ApJ}           % Astrophysical Journal
\newcommand{\mnras}{MNRAS}       % Monthly Notices of the RAS

\graphicspath{{Figures/}}

\title{The Kinematic Morphology-Density Relation from the SAMI Pilot Survey.}

\author[Fogarty, L.~M.~R. ]{L.~M.~R. Fogarty$^{1,2}$ and the SAMI Galaxy Survey Team$^3$}

\affiliation{$^1$Sydney Institute for Astronomy (SIfA), School of Physics, The University of Sydney, NSW 2006, Australia\\
$^2$ARC Centre of Excellence for All-sky Astrophysics (CAASTRO) \\
email: {\tt l.fogarty@physics.usyd.edu.au}\\
$^3$ Full list of team members is available at {\tt http://sami-survey.org/members}}

\pubyear{2014}
\volume{309}
\jname{Galaxies in 3D across the Universe}
\editors{B. L. Ziegler, F. Combes, H. Dannerbauer, M. Verdugo, eds.}

\begin{document}

\maketitle

\begin{abstract}

We present the kinematic morphology-density relation in three galaxy clusters, Abell 85, 168 and 2399, using data from the SAMI Pilot Survey. We classify the early-type galaxies in our sample as fast or slow rotators (FRs/SRs) according to a measured proxy for their projected specific stellar angular momentum. We find each cluster contains both fast and slow rotators with and average fraction of SRs in the sample of $f_{SR}=0.15\pm$0.04. We investigate this fraction within each cluster as a function of local projected galaxy density. For Abell 85 we find that $f_{SR}$ increases at high local density but for Abell 168 and 2399 this trend is not seen. We find SRs not just at the centres of our clusters but also on the outskirts and hypothesise that these SRs may have formed in group environments eventually accreted to the larger cluster.


\end{abstract}

\firstsection
\section{Introduction}

%-Intro to kinematic morphology density relation.
%-Intro to SAMI and SAMI Galaxy Survey.

The morphology-density relation \citep{Dressler1980} shows that the number density of early-type galaxies (ETGs, typically classified morphologically as lenticular (S0) and elliptical (E) galaxies) increases as one proceeds from low density to high density environments, at the expense of spiral galaxies whose number density declines comparatively. This robust result holds for galaxy clusters with diverse properties, including relaxed and unrelaxed systems, high and low mass clusters and some clusters with strong X-ray emission.

In recent decades the traditional definition of ETGs has been refined to include information gathered using 3D data from integral field spectroscopy (IFS). Building on results from the SAURON Survey \citep{deZeeuw2002}, the ATLAS$^{\rm{3D}}$ Survey \citep{Cappellari2011a} studied a volume limited sample of 260 ETGs out to 42\,Mpc distance. The unprecedented detail of these 3D observations led to a re-classification of ETGs from a morphological basis to a kinematic one. The new framework \citep{Emsellem2007, Emsellem2011} uses the kinematics of ETGs to classify them as fast rotators (FRs), disk-supported systems with high specific stellar angular momentum, or slow rotators (SRs), dispersion-supported systems with low specific stellar angular momentum. Analogous to the morphology-density relation \citet{Cappellari2011b} found that the fraction of SRs in the galaxy population increases in the highest density environments. This is the kinematic morphology-density relation. This relation can be used to study the formation and evolution of ETGs but requires large samples of integral field spectroscopy (IFS) data.

The Sydney-AAO Multi-Object Integral Field Spectrograph \citep[SAMI,][]{Croom2012} is a fibre-based multi-object spectrograph on the AAT capable of observing 13 galaxies at one time, thus increasing IFS survey efficiency by an order of magnitude. With SAMI we have completed the SAMI Pilot Survey, presented here. We observed 106 galaxies across three galaxy clusters (Abell 85, 168 and 2399) with IFS data for each.

\begin{figure}
\centering
\includegraphics[width=0.8\columnwidth]{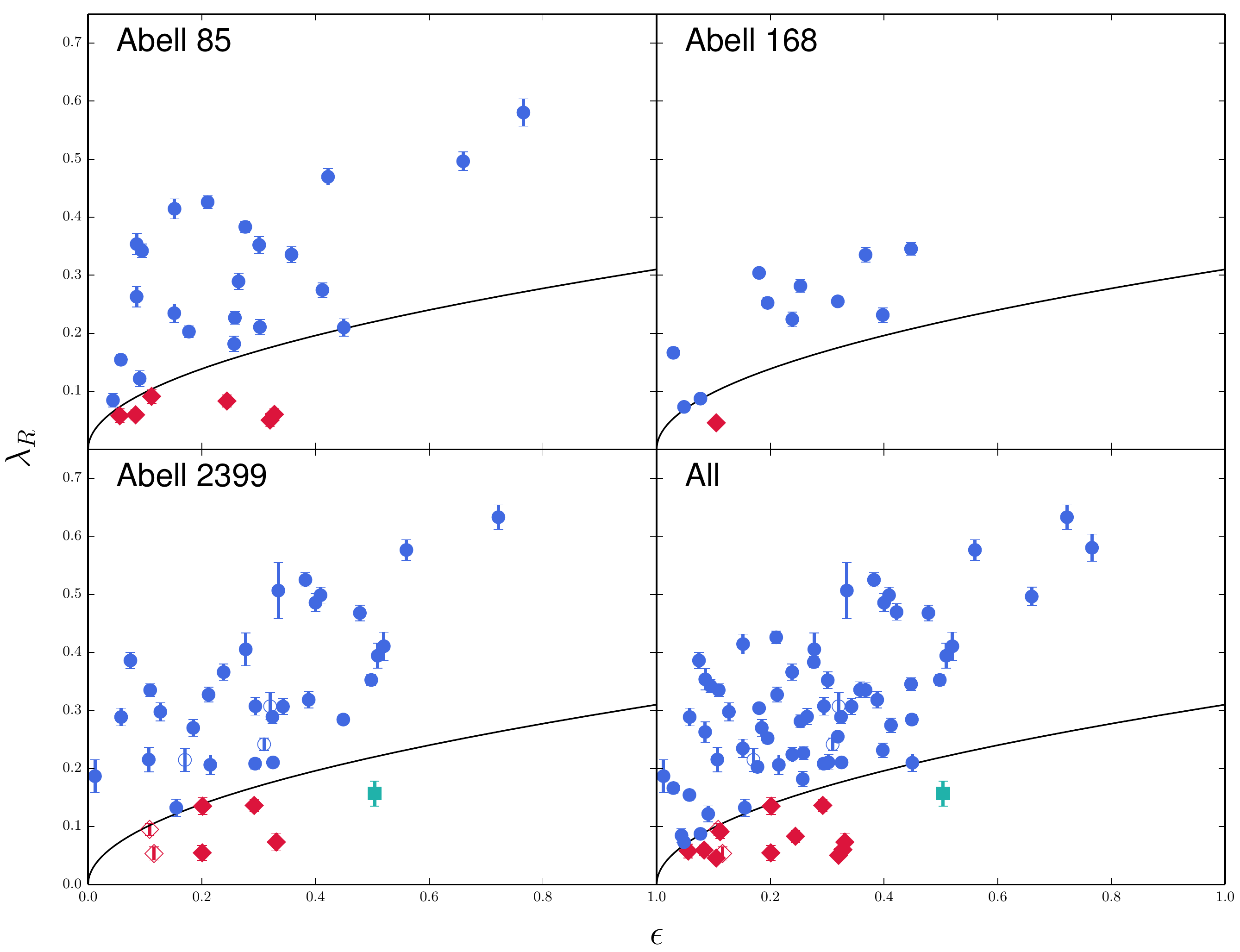} 
\caption{The $\lambda_{R}-\epsilon$ diagram for Abell 85 (top left), 168 (top right), 2399 (bottom left) and the entire SAMI Pilot Survey sample (bottom right). The blue circles are FRs and the red diamonds SRs. The turquoise square point indicates a double sigma galaxy - a disk supported system interloping into SR space. As such it is excluded from our statistics.}
\label{fig:lambda_r}
\end{figure}

\section{Galaxy Data and Kinematic Classification}

The SAMI Pilot Survey was carried out over 10 nights at the AAT in September and October 2012. Full details of the observations are given in \citet{Fogarty2014} and Fogarty et al. in prep.. A total of 106 galaxies of all morphological types were successfully observed. The majority of these were cluster member galaxies, with a handful of foreground and background objects. Once our observations were complete we morphologically classified our sample by eye, in order to isolate the ETGs. In total 79 ETGs were observed.

SAMI feeds the double-beamed AAOmega spectrograph \citep{Smith2004} which is fully configurable. For all SAMI Pilot Survey observations we used the 5700\,\AA\, dichroic with a mid-resolution 580V grating in the blue arm (covering the wavelength range $3700-5700$\,\AA\, with $\rm{R}=1700$) and a high-resolution grating in the red arm (covering the wavelength range $6200-7300$\,\AA\, with $\rm{R}=4500$). Each observation produces two fully-calibrated data cubes with variance and covariance information fully propagated through the reduction process. The data reduction procedure for SAMI is described fully in \citet{Allen2014} and \citet{Sharp2014}. 

\begin{figure}
\centering
\includegraphics[width=0.8\columnwidth]{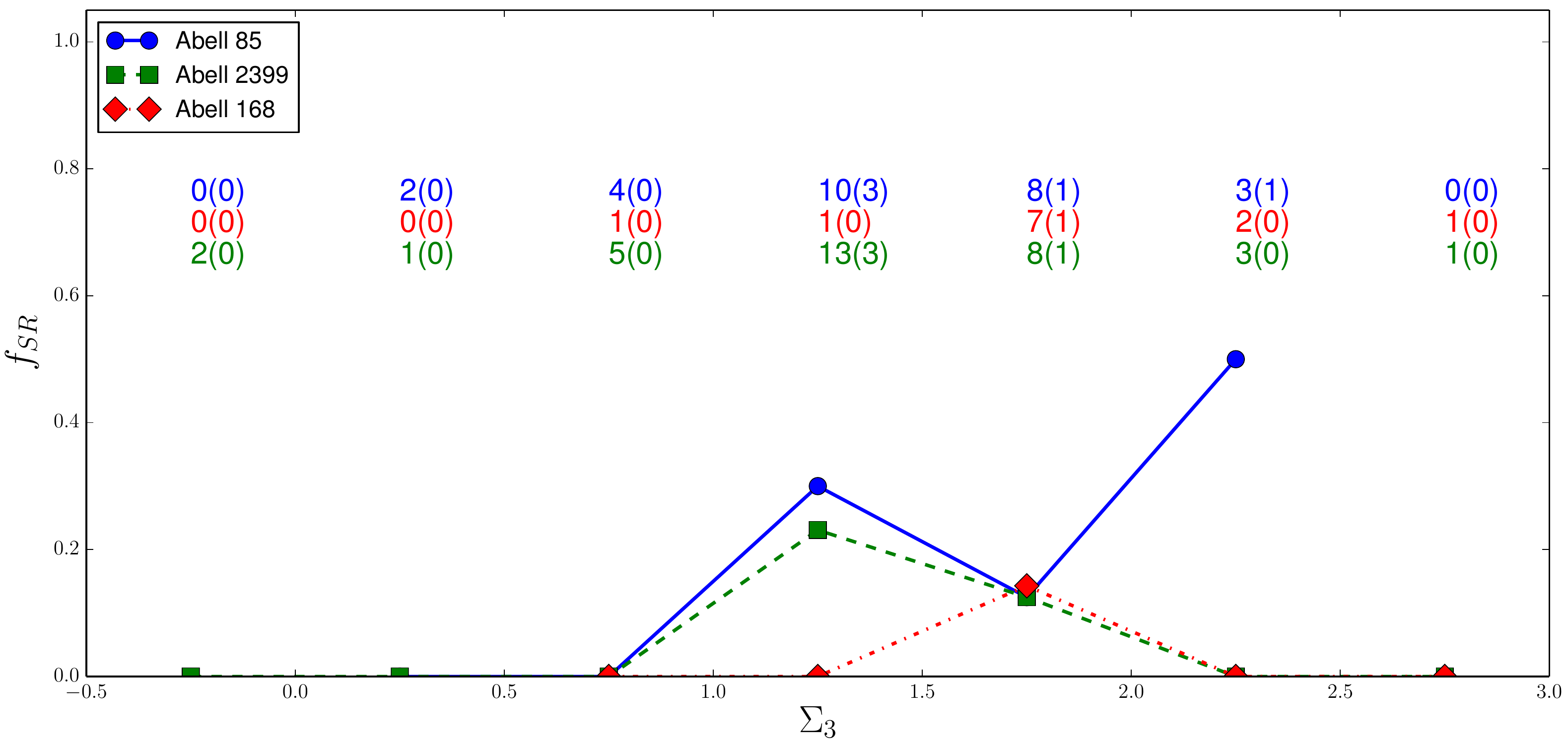} 
\caption{The kinematic morphology-density relation for Abell 85 (blue circles, solid line), 168 (red diamonds, dash-dot line) and 2399 (green squares, dashed line). The number of galaxies (SRs) is given for each bin over the bin.}
\label{fig:morph_dens}
\end{figure}

The blue SAMI data cubes cover many important stellar absorption features and are used to derive stellar velocity and velocity dispersion maps for each galaxy. We use the penalised pixel fitting technique of \citet{CappellariEmsellem2004} and the MILES empirical spectral templates \citep{SanchezBlazquez2006}. The kinematic maps are then used to calculate $\lambda_{R}$ \citep{Emsellem2007, Emsellem2011}, a proxy for the projected specific stellar angular momentum for each galaxy. The classification system of \citet{Emsellem2011} is used whereby a galaxy is considered a SR if $\lambda_{R}<k\epsilon$ where the value of $k$ is determined by the fiducial radius within which one measures $\lambda_{R}$ and $\epsilon$ (galaxy ellipticity). Since our galaxies span a range of sizes we use three fiducial radii, Re/2, Re and 2Re for large, intermediate and small objects respectively. This does not bias our classification, as shown in \citet{Fogarty2014}. 

Figure \ref{fig:lambda_r} shows the $\lambda_{R}-\epsilon$ diagram for all three of our clusters and the entire sample combined. We find SRs in all three clusters studied and the fraction of SRs in the ETG population is $0.21\pm0.08$, $0.08\pm0.08$ and $0.12\pm0.06$ for Abell 85, 168 and 2399 respectively. The overall fraction for the entire sample is $f_{SR}=0.15\pm$0.04. This is consistent with previous studies, which find an overall value of $f_{SR}\sim0.15$ across a range of average global environments \citep{DEugenio2013, Houghton2013}. 
\section{The Kinematic Morphology-Density Relation}

We examine the value of $f_{SR}$ as a function of local projected galaxy density. This is parametrised by the third nearest neighbour surface density, $\Sigma_3$, and the method by which $\Sigma_3$ is calculated is described in \citet{Fogarty2014}. Figure \ref{fig:morph_dens} shows the kinematic morphology-density relation for Abell 85 (blue circles), 168 (red diamonds) and 2399 (green squares). For Abell 168 our sampling with the SAMI Pilot sample is poor so it is not possible to draw any robust conclusions for this cluster. For Abell 85 we see the expected trend such that $f_{SR}$ increases in the densest part of the cluster. However, this trend is not seen in Abell 2399 where the two central BCG candidate galaxies are in fact FRs. In both Abell 85 and 2399 we also see a high $f_{SR}$ at low to intermediate densities within each cluster, indicating that SRs are not only found in the centres of clusters but also on their outskirts.

The roughly constant value of $f_{SR}$ we find (consistent with previous studies, e.g. \citet{Houghton2013}) in all three clusters suggests that the formation mechanism for SRs is equally efficient in environments as diverse as the field, galaxy groups and low-mass to high-mass clusters. However, SRs are preferentially found at cluster centres in many previous studies \citep{Cappellari2011b, DEugenio2013, Houghton2013}. Here we see a more complex picture with one cluster obeying the expected kinematic morphology-density relation but with evidence as well for in-falling SRs. One possible explanation is that SRs could form preferentially as central galaxies in a halo, such as a galaxy group. These halos then grow into large clusters dominated by SRs or are eventually accreted to already existing clusters. The large SRs then make their way to the centres of their host clusters by dynamical friction. We hypothesise that the SRs on the outskirts of Abell 85 and 2399 were formed in groups and are joining their host clusters through merging processes. This is supported by the fact that all three of the clusters studied for the SAMI Pilot Survey are known cluster-cluster mergers.

\section{Conclusions}

We have kinematically classified the 79 ETGs of the SAMI Pilot Survey of three galaxy clusters, Abell 85, 168 and 2399. We calculated a SR fraction of $0.21\pm0.08$, $0.08\pm0.08$ and $0.12\pm0.06$ for each cluster respectively with an overall fraction of $f_{SR}=0.15\pm$0.04 for the entire sample. This is consistent with previous work in this area and the hypothesis that the SR formation mechanism is equally efficient across many different global environments from the field to high-mass clusters.

In addition we examined the kinematic morphology-density relation for all three clusters. We find that although in Abell 85 we see the expected trend such that $f_{SR}$ increases in the densest part of the cluster we also see significant evidence for the existence of SRs in low and intermediate density regions of both Abell 85 and 2399. We hypothesise that these SRs could have formed as centrals in groups and later fallen into the cluster as SRs.

% Overall fraction consistent with literature
% SRs found across many densities
% SR formation in groups

\section*{Acknowledgements}

\noindent The SAMI Galaxy Survey is based on observations made at the Anglo-Australian Telescope. The Sydney-AAO Multi-object Integral field spectrograph (SAMI) was developed jointly by the University of Sydney and the Australian Astronomical Observatory. The SAMI input catalogue is based on data taken from the Sloan Digital Sky Survey, the GAMA Survey and the VST ATLAS Survey. The SAMI Galaxy Survey is funded by the Australian Research Council Centre of Excellence for All-sky Astrophysics (CAASTRO), through project number CE110001020, and other participating institutions. The SAMI Galaxy Survey website is http://sami-survey.org/.

\end{document}